\documentclass[preprint,longbibliography,aps,pra,superscriptaddress,showpacs,floatfix]{revtex4-1}
\usepackage{graphics,epsfig,graphicx}
\usepackage{amsbsy}
\usepackage{amsmath}
\usepackage{bm}
\usepackage{amsfonts}
\usepackage{cancel}
\usepackage{multirow}
\usepackage{svn-multi}
\begin{document}

\title{The Fate of excited state of $^4\text{He}$ }

\author{M. Gattobigio}
\affiliation{
 Universit\'e C\^ote d'Azur, CNRS, Institut  de  Physique  de  Nice,
17 rue Julien Laupr\^etre, 06200 Nice, France}
\author{A. Kievsky} 
\affiliation{Istituto Nazionale di Fisica Nucleare, Largo Pontecorvo 3, 56100 Pisa, Italy}
\begin{abstract}
  We investigate the properties of the excited state of $^4\mathrm{He}$, 
  $^4\mathrm{He}^*$, within the framework of Efimov physics and its connection
  to the unitary point of the nuclear interaction. We explore two different
  approaches to track the trajectory of $^4\mathrm{He}^*$ as it crosses the
  $^3\mathrm{H}$+p threshold and potentially becomes a resonant state. The first
  approach involves an analytical continuation of the energy with respect to the
  Coulomb coupling, while the second approach introduces an artificial four-body
  force that it is gradually released. By utilizing Pad\'e approximants and
  extrapolation techniques, we estimate the energy and width of the resonance.
  Our results suggest a central energy value of $E_R=0.060(3)$ MeV and a width
  of $\Gamma/2=0.036(6)$ MeV using the Coulomb analysis, and $E_R=0.068(1)$ MeV
  and $\Gamma/2=0.007(5)$ MeV with the four-body force analysis. Interestingly,
  these results are consistent with calculations based on {\it ab-initio}
  nuclear interactions but differ from the accepted values of the $0^+$
  resonance energy and width. This highlights the challenges in accurately
  determining the properties of resonant states in light nuclei and calls for
  further investigations and refinements in theoretical approaches.
\end{abstract}
\maketitle
\section{Introduction}

Universal properties can be identified in light nuclear systems in association
with what is called Efimov physics~\cite{gattobigio:2019_Phys.Rev.C,
kievsky:2018_J.Phys.Conf.Ser.,kievsky:2016_Few-BodySyst}.  This is related to
the shallow character of the neutron-proton (np) bound state in the $J^\pi=1^+$
channel, the deuteron, and the shallow character of the virtual nn and np states
in the $0^+$ channel. As a consequence the two-nucleon singlet and triplet
scattering lengths are much larger than the typical interaction length of the
nuclear force, approximately 1.5 fm. These properties locate the two-nucleon
system close to the unitary point, a point characterized by the energies of
those states set to zero. The appearance of a soft scale allows for an effective
description of the nuclear states as discussed in
Ref.~\cite{hammer:2020_Rev.Mod.Phys.}.  Moreover, the large values of the
two-body scattering lengths introduce correlations that propagate to the three-
and four-nucleon systems with the most remarkable property known as the Efimov
effect~\cite{efimov:1971_Sov.J.Nucl.Phys.}.

Efimov physics is primarily associated with equal boson systems, while nuclei
are composed of fermions. However, the crucial point in applying Efimov physics
to identical particles lies in the symmetry of the spatial wave function. The
dominant portion of the spatial wave functions of the $1/2^+$ state of the
triton, $^3$H, of the $^3$He, and of the $0^+$ state of $^4$He, is symmetric, indicating
that these states belong to this sector of physics in which universal
properties can be observed.  For example the energy ratio $E_{^4{\rm
He}}/E_{^3{\rm He}}=3.7$ is close to the same ratio of 4.6 calculated in the
zero-range limit~\cite{deltuva:2013_Few-BodySyst} at the unitary point, or to
4.4, the ratio of the bosonic tetramer and trimer energies of helium 
atoms~\cite{hiyama:2012_Phys.Rev.A}. When Coulomb effects are neglected the
nuclear ratio increases up to 3.9~\cite{gattobigio:2019_Phys.Rev.C}, however
finite-range effects still have to be evaluated~\cite{kievsky:2015_Phys.Rev.A}.

In the four-body sector, the Efimov scenario predicts two bound
$0^+$ states, the ground state and one excited state, with the latter indicated
as $^4\text{He}^*$, lying just below the three-body threshold given by 
$^3$H~\cite{platter:2004_Phys.Rev.A,hadizadeh:2012_Phys.Rev.A,%
gattobigio:2012_Phys.Rev.A,alvarez-rodriguez:2016_Phys.Rev.A}.
The spectrum of $^4$He does not show such an
excited state. However, there exists a resonant state having the same quantum
numbers as the ground state.  In this paper, we will explore the possibility
that the $0^+$ resonance of $^4$He is a remnant of the four-body excited state
$^4\text{He}^*$ pushed to the continuum by the presence of the Coulomb
interaction.

The $0^+$ resonance of $^4$He has been experimentally measured, and its position
has been analyzed using the $R$-matrix approach~\cite{tilley:1992_Nucl.Phys.A}.
The result of this analysis is that the resonance has an
energy with respect to the $^3$H threshold 

\begin{equation}
  E_R + i\,\Gamma/2 = 0.39\,\text{MeV} + i\,0.50/{2} \,\text{MeV}\,.
  \label{eq:experimentalResonance}
\end{equation}

In the literature, there are several calculations of this resonance using
different potentials such as the Minnesota potential and realistic
potentials~\cite{viviani:2018_Few-BodySyst,aoyama:2016_Prog.Theor.Exp.Phys.},
but these calculations do not consistently reproduce the experimental position
of the resonance.
For instance, calculations with Minnesota potential give $E_R \simeq
0.07$~MeV~\cite{viviani:2018_Few-BodySyst,aoyama:2016_Prog.Theor.Exp.Phys.},
while realistic potentials predict $E_R\simeq 0.08(1)$~MeV~\cite{viviani:2018_Few-BodySyst}.  Additionally, a
significant source of uncertainty is the width of the resonance, which can be
affected by the choice of potential and particularly by the inclusion of a
three-body force~\cite{viviani:2018_Few-BodySyst,hofmann:2008_Phys.Rev.C}.

In this paper, we use a simple spin-isospin-dependent Gaussian potential to
investigate the resonance. This potential captures the fundamental principles of
Efimov physics and provides a Gaussian characterization of systems within the
universality
window~\cite{gattobigio:2019_Phys.Rev.C,kievsky:2021_Annu.Rev.Nucl.Part.Sci.,
kievsky:2020_Phys.Rev.A,deltuva:2020_Phys.Rev.C}.
Without considering the Coulomb potential, the Gaussian potential predicts two bound states
for the four-body system. As the Coulomb interaction is gradually turned on, it
pushes the excited state beyond the tritium-proton ($^3$H+p) threshold.
We will explore this trajectory using two distinct approaches, both of which
rely on the technique of analytic continuation in the coupling constant
(ACCC)~\cite{kukulin:1989_}. The first approach involves applying the ACCC to
the Coulomb interaction itself, while the second approach involves applying the
ACCC to a four-body force specifically introduced to relocate the resonance into
the bound state region. By employing the ACCC to the four-body force for the
newly formed bound state, we can extrapolate its energy value in the continuum region.

Based on our investigations using this simple model, our findings suggest that
the excited state transforms into a resonant state with an energy that aligns
with the results obtained in previous studies.
We conclude that the $0^+$ resonance of $^4$He is, in fact, the excited state of
the Efimov characterization of the light-nuclei sector of nuclear physics, which
has been pushed beyond the $^3$H+p threshold by the Coulomb interaction. 
There is a general belief that the discrepancy between the
numerical calculations based on realistic potential and experimental results can
be due to the limitations of the $R$-matrix analysis of the resonance.

\section{Gaussian characterization}

Gaussian potentials have demonstrated their suitability in describing few-body
systems within the universal window~\cite{kievsky:2016_Few-BodySyst,
kievsky:2021_Annu.Rev.Nucl.Part.Sci.}.
Specifically, they provide an excellent description of the spectrum of light
nuclei~\cite{gattobigio:2019_Phys.Rev.C}. Moreover, Gaussian potentials have
recently been employed to extract the Coulomb-free $^1S_0$ proton-proton
scattering length in an almost model-independent manner~\cite{tumino:2023_Commun.Phys.}.

In our model, we set the nucleon mass such that 
$\hbar^2/m = 41.47~\text{MeV}\,\,\text{fm}^2$, 
and the interaction to be a 
two-body Gaussian potential that only acts on the 
spin/isospin $(S/T)$ channels $S=0$, $T=1$, and $S=1$, $T=0$
\begin{equation}
  V(r)  =  V_{01}\,e^{-(r/r_0)^2} {\cal P}_{01}
        +  V_{10}\,e^{-(r/r_0)^2} {\cal P}_{10}\,,
  \label{eq:twoBody}
\end{equation}
where we have introduced the spin-isospin projectors $ {\cal P}_{ST}$. 

This choice results in the dominance of the s-wave channel, consistent with
Efimov physics. The decision to exclude a p-wave contribution is motivated by
its limited role in the current context.
We set the value of $r_0=1.65$~fm, and choose the potential
strengths as $V_{01} = -37.9$~MeV and $V_{10}=-60.575$~MeV.
These values are selected to accurately reproduce the binding energy of the
deuteron and provide the best description of the single and triplet scattering
lengths and effective ranges.
Moreover, we introduce a three-body force 
\begin{equation}
  W(\rho_3)  = W_0\,e^{- (r_{12}^2+r_{13}^2+r_{23}^2)/R_3^2} \,,
  \label{eq:threeBody}
\end{equation}
which is a function of the hyperradius $\rho_3$, and 
whose parameters are tuned to give the best description of $^{3}$H energy, and,
once Coulomb interaction has been switched on, also the best description of
$^{3}$He and $^{4}$He energies.  With $W_0=7.6044$~MeV and $R_3=3.035$~fm, we
obtain $E_{^3\text{H}} = -8.482$~MeV, $E_{^3\text{He}}=-7.716$~MeV, and  
$E_{^4\text{He}}=-28.305$~MeV. A summary of the parameters we used and of the energies
we obtained, together with the experimental ones, are given in
Table~\ref{tab:physicalPoint}. A more complete discussion of the
choice of the parameters can be found in Ref.~\cite{gattobigio:2019_Phys.Rev.C}.

\begin{table}
  \caption{Gaussian potential parameters and corresponding
	$^3$H, $^3$He and $^4$He ground-state binding energies.
  For the sake of comparison, the experimental values are shown too.}
  \label{tab:physicalPoint}
\begin{tabular} {@{}c c c c c c c c c c@{}}
\hline\hline
$V_{01}$(MeV) & $V_{10}$(MeV) & $r_0$(fm) & $W_{0}$(MeV) & $R_3$(fm)  &  $^3$H(MeV) & $^3$He(MeV) & $^4$He(MeV)   \\ 
\hline
-37.9 & -60.575 & 1.63 & 7.6044 & 3.035 & -8.482 & -7.716 &  -28.305 \\
\cline{6-8}
\multicolumn{5}{c}{Experimental Values} & -8.482 & -7.718  & -28.296& \\
 \hline
\hline
\end{tabular}
\end{table}

In the hypothetical scenario where the Coulomb interaction is not taken into
account, the four-nucleon system would exhibit two bound states: one deep and
one shallow, associated with the three-body ground state. This two-level
structure shows universal behavior, the energy ratios remain
independent of the specific system being 
considered~\cite{deltuva:2013_Few-BodySyst}. Our calculations
provide evidence that nuclear physics also falls within the realm of the
universal Efimov window. However, the inclusion of the Coulomb interaction
transforms the excited state of the four-body system into a resonance located
above the $^3$H+p threshold. 
It should be noticed that in the three-nucleon
system something similar can be observed. Tuning the two-body gaussian
potential strengths to make the singlet and triplet scattering lengths diverge,
the three-nucleon system will show the infinite set of Efimov states.
Tuning back the strengths to match their physical values again, 
the tower of Efimov states disappear one by one and, at the physics point, one
state will survive, the triton. It is well known that the triton has an excited
virtual state and, similarly to the analysis doing here, it can be traced from 
the last Efimov state that crossed the neutron-deuteron
threshold~\cite{rupak:2019_Phys.Lett.B,deltuva:2020_Phys.Rev.C}.

\begin{figure}
  \begin{center} 
    \includegraphics[width=0.7\linewidth]{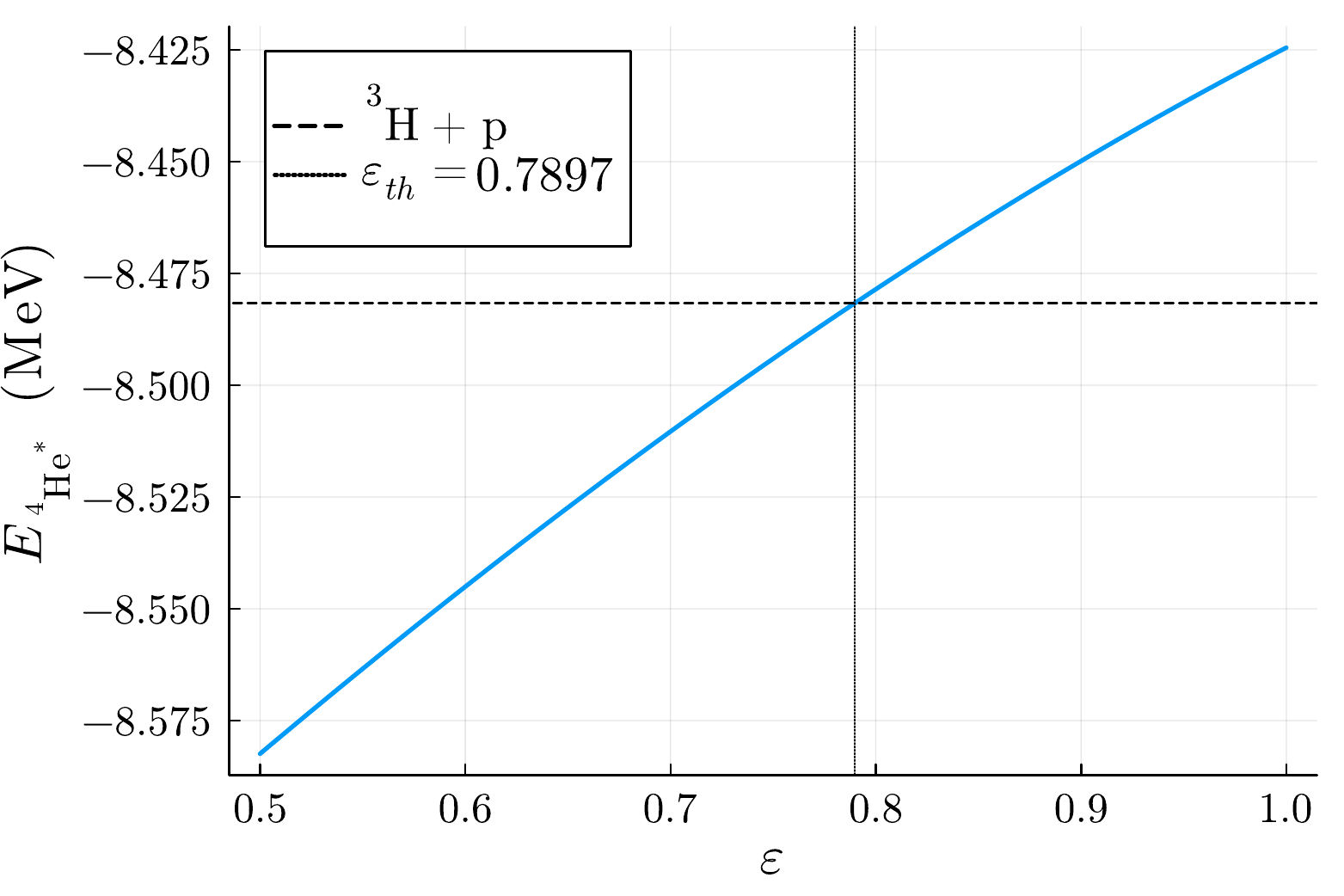}
  \end{center}
  \caption{The energy of the excited state $^4\mathrm{He}^*$ as a function of the 
  parameter $\epsilon$ controlling the Coulomb interaction. 
  The horizontal dashed line corresponds to the $^3$H+p threshold. 
  The excited state goes through the threshold at 
  $\epsilon_{th}=0.7897$, which is indicated by the vertical dotted line.}
  \label{fig:threshold}
\end{figure}

Here, we track the change on the four- and three-body states as the Coulomb interaction is gradually
introduced through the use of a parameter $\epsilon$ which 
multiplies the Coulomb potential 
\begin{equation}
	V_c(r_{pp})=\epsilon \frac{e^2}{r_{pp}} \,,
\end{equation}
where $r_{pp}$ is the distance between two protons.
For $\epsilon=0$, that means no
Coulomb interaction, the two three-nucleon systems, $^3$He and
$^3$H, are degenerate in energy (in the present analysis we are disregarding small charge symmetry
breaking effects). Moreover, two four-body states appear that can be 
identified as the ground and an excited state of  $^{4}$He. 
Switching on the Coulomb interaction the degeneracy between the $^3$H
and $^3$He is lifted and the energies of the ground and excited state 
of $^4$He begin to shift; in particular, for $\epsilon\approx 0.78$, the $^4\text{He}$ excited
state goes behind the $^3$H+p threshold and becomes a resonance. 
In Fig.~\ref{fig:threshold}, where the energy of the excited state
$E_{^4\mathrm{He}^*}$ is given as a function of $\epsilon$,
we see that the threshold value is $\epsilon_{th}=0.7897$.

\section{Extrapolation in the continuum - ACCC}
In this section we use the ACCC~\cite{kukulin:1989_} technique in order to follow the energy
of $^4\text{He}^*$ behind the threshold and thus in the complex plane.  This
extrapolation procedure can be very subtle; in fact, once a state goes through a
threshold, the analytical behavior of the energy as a function of the coupling
constant, is non trivial~\cite{bender:1993_Phys.Lett.Aa}.  

To track the $^4\text{He}^*$ state, we will employ two distinct procedures. The
first procedure involves analytic continuation in the $\epsilon$ parameter,
which corresponds to the Coulomb interaction. During the gradual switching on, 
the $^3$H+p threshold energy remains constant. We will follow the
methodology outlined in~\cite{kukulin:1989_}, utilizing Pad\'e approximants to
describe the energy of $^4\text{He}^*$ as a function of $\epsilon$ and
analytically continue it up to $\epsilon=1$.

The second procedure relies in the introduction of a four-body 
force, more in the spirit of Ref.~\cite{kukulin:1989_}. If the
four-body force has enough attraction, the resonance is brought back
to the physical sheet, that means it becomes a bound state, without
changing the value of the thresholds determined by the three-nucleon systems. 
Subsequently, the force is gradually weakened until it reaches a critical value,
allowing us to track the state below the threshold using the analytic
continuation provided by the Pad\'e approximant.  
Remarkably, the results obtained for the energy of the resonance using both
methods are found to be compatible with each other.

For a two-body system the ACCC theory is well established, while for the 
multi-body problem the theory is absent. However, there are several studies
suggesting that the behavior of the energy as a function of the coupling
constant is similar to the two-body case if the decay channel is two-body, or
like a two-body case with an effective barrier if the decay channel is
multi-body~\cite{tanaka:1999_Phys.Rev.C}.

Let's introduce the complex-valued momentum for the state of $N$ particles with
energy $E_N$ with respect to the threshold $E_{N-1}$ of the subsystem with $N-1$
particles 
\begin{equation} 
  k(g) = \sqrt{\frac{E_N-E_{N-1}}{\hbar^2/m}} \,,
  \label{eq:momentum}
\end{equation} 
that depends on a coupling constant $g$ used to move the $N$-body energy $E_N$.
In our context, the coupling constant $g$ corresponds to $\epsilon$ in the case
we are using the method in correspondence with the
Coulomb interaction and represents the strength of the four-body potential in
the case the method is used introducing a four-body force. 
When the $N$-particle state lies below the $N-1$ threshold
it is bound, and the momentum $k(g)$ , as given by Eq.~(\ref{eq:momentum}), lies on the
positive imaginary axis. In the following discussion, we assume that the
threshold is attained at the specific value $g_{th}$ of the coupling constant.
Moreover, we consider the system to be below the threshold for $g<g_{th}$.

In the vicinity of the threshold, the theory for the two-body system with
short-range interactions predicts~\cite{kukulin:1989_}

\begin{equation}
  k(g) \sim \left\{
    \begin{matrix}
    i (g_{th}-g) & \ell = 0  \\[0.1cm]
      i\sqrt{g_{th}-g} & \ell > 0 \,,
    \end{matrix}\right.
    \label{eq:trend}
\end{equation}
where $\ell$ is the relative angular momentum between the two
particles. 

In all our calculations, we consistently observe that $k(g) \sim i
\sqrt{g_{th}-g}$ (see Figs.~\ref{fig:behaviour} and \ref{fig:behaviour4B}). 
This behavior indicates that the threshold coupling constant
$g_{th}$ coincides with the branch point of the momentum $k(g)$, which is a
non-analytic function of the coupling constant~\cite{kukulin:1989_}. To track
the physical branch, we introduce the variable $\zeta = \sqrt{g_{th}-g}$ and
parametrize the momentum using a
Pad\'e approximant of order $[N,N]$
\begin{equation}
  k(\zeta) = i\,\frac{a_0+a_1 \zeta+\cdots+a_N \zeta^N}{1+b_1 \zeta + \cdots + b_N \zeta^N} \,.
  \label{eq:pade}
\end{equation}

The coefficients $a_0, \dots, a_N$ and $b_1, \dots, b_N$ are determined within
the region where the state is bound. Once these coefficients are obtained, we
extrapolate the value of $k(\zeta)$ to the physical value $g_{ph}$ of the
coupling constant. At this point, the momentum typically becomes a complex
number, enabling us to extract the complex energy
\begin{equation}
  E_N(g_{ph}) = \frac{\hbar^2 k^2(g_{ph})}{m}\,.
\end{equation} 
The physical value $g_{ph}$ depends on the specific case. In the Coulomb case,
the physical value of the coupling constant is $\epsilon\equiv g_{ph} = 1$, while in the
four-body force case, the physical value is $g_{ph} = 0$.

\subsection{Coulomb ACCC}

\begin{figure}
  \begin{center} 
    \includegraphics[width=0.7\linewidth]{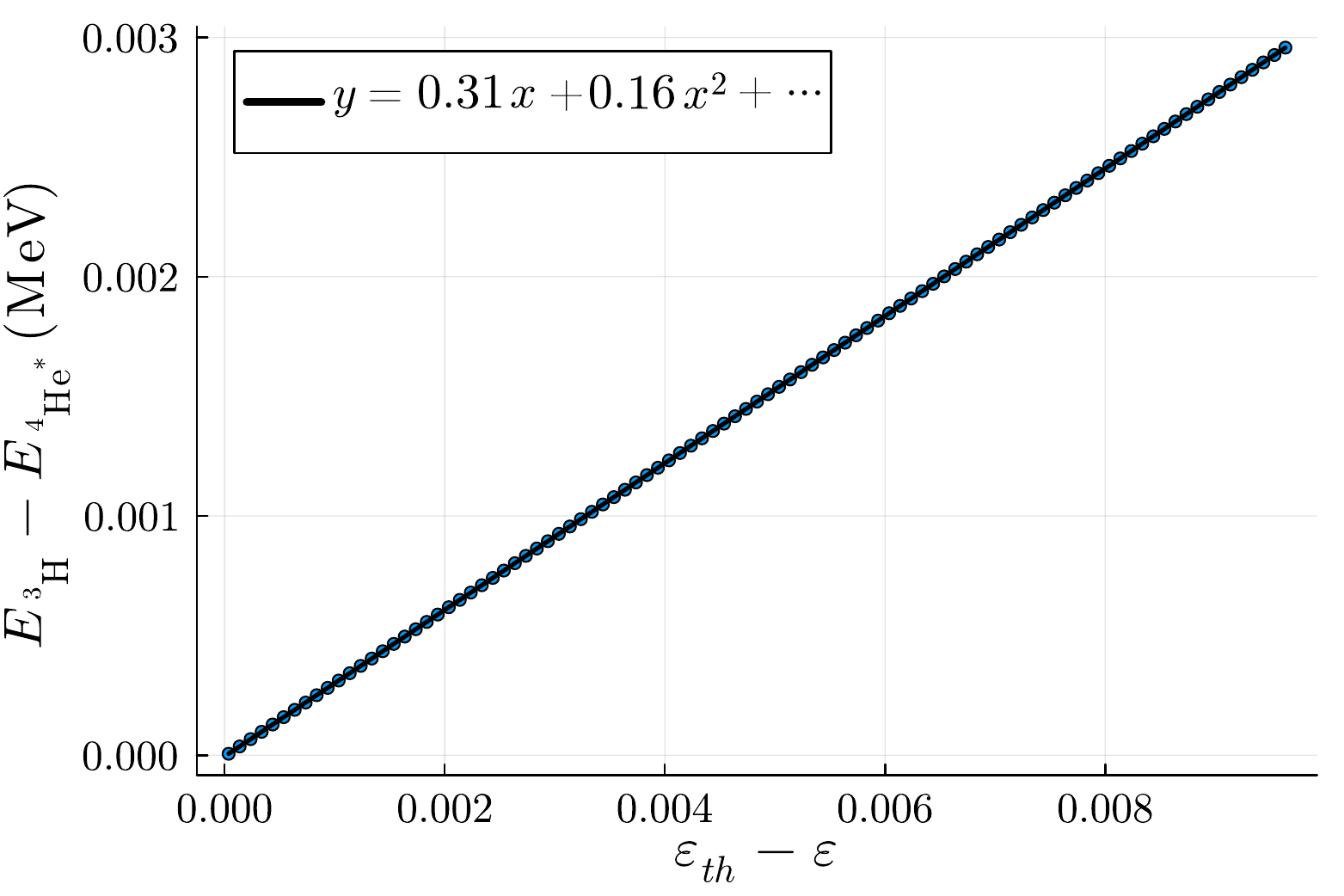}
  \end{center}
  \caption{
    The energy difference between the four-body excited state and the $^3$H+p
    threshold as a function of $\epsilon_{th}-\epsilon$. A polynomial fit
    reveals a dominant linear behavior, implying $k(\epsilon) \sim i
    \sqrt{\epsilon_{th}-\epsilon}$.}
  \label{fig:behaviour}
\end{figure}

In the case of Coulomb interaction, the coupling constant is denoted as
$\epsilon$ and varies between zero (no interaction) and one (the physical case).
The threshold value, determined from Fig.\ref{fig:threshold}, is $\epsilon_{th}
= 0.7897$. Obtaining the value of $\epsilon_{th}$ is a critical step in our
analysis. We utilized a correlated-Gaussian basis set, optimized using the
stochastic variational method~\cite{suzuki:1998_} for $\epsilon = 0.78$. This
optimized basis set was then employed to determine the threshold.
In Fig.~\ref{fig:behaviour}, we report the energy difference between the
three-body threshold and the energy of the four-body excited state as a function
of $\epsilon_{th}-\epsilon$. We clearly observe a dominant linear behavior.

We can now proceed with the fitting of the momentum using the Pad\'e approximant
given by Eq.(\ref{eq:pade}), while varying the degree $N$ of the polynomials.
Once we have determined the Pad\'e approximant, we can calculate the value of the momentum
at $\epsilon = 1$, allowing us to extrapolate the resonance energy. The results
are presented in Table~\ref{tab:resonanceCoulomb}. We observe that there is a
range of values for $N$ where the extrapolation remains quite stable. We obtain
$E_R = 0.060(3)$ MeV for the center of the resonance and $\Gamma/2 = 0.036(6)$
MeV for the width. 
\begin{table}
  \caption{Extrapolated values of the momentum and of the energy of the 
  resonance as a function of the degree of the Pad\'e approximant
  for Coulomb potential.}
  \label{tab:resonanceCoulomb}
\begin{tabular} {@{}c c c c c@{}}
\hline\hline
$N$ & Im($k$)(fm$^{-1}$) & Re(k)(fm$^{-1}$)  & $E_R$ (MeV) & $\Gamma/2$ (MeV)\\ 
\hline
4 &  0.039 &-0.0011 &0.063 & 0.0036 \\
5 &  0.038 &-0.0014 &0.060 & 0.0044 \\
6 &  0.038 &-0.0012 &0.060 & 0.0038 \\
7 &  0.038 &-0.0010 &0.060 & 0.0032 \\ 
8 &  0.037 &-0.0010 &0.057 & 0.0031 \\
9 &  0.038 &-0.0012 &0.060 & 0.0038  \\
\hline
\hline
\end{tabular}
\end{table}

\subsection{Four-body force ACCC}
\begin{figure}
  \begin{center} 
    \includegraphics[width=0.7\linewidth]{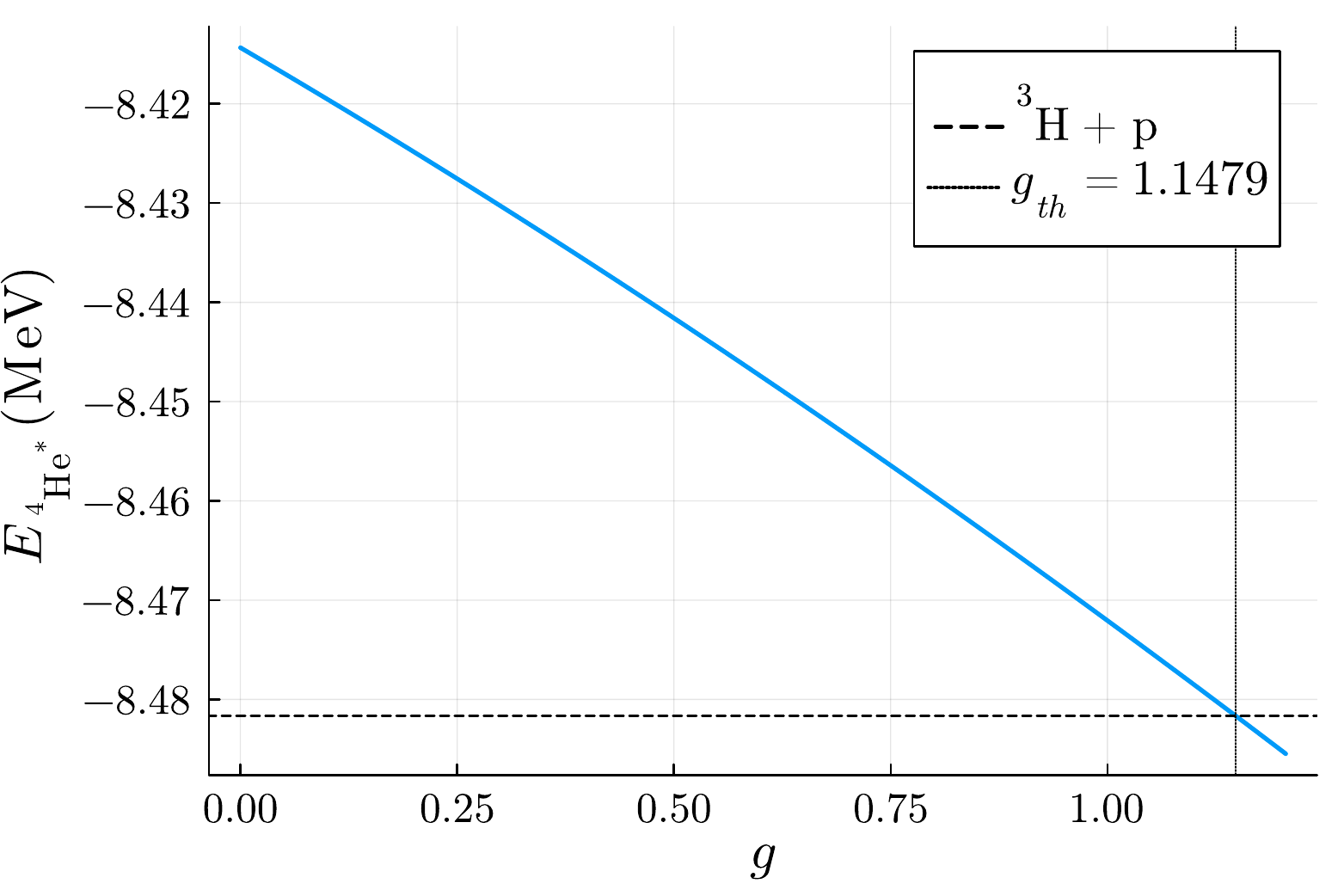}
  \end{center}
  \caption{The energy of the the 
  the excited state $^4\mathrm{He}^*$ as a function of the four-body strength $g$.
  The horizontal dashed line corresponds to the $^3$H+p threshold. 
  The excited state goes through the threshold at 
  $g_{th}=1.1479$, which is indicated by the vertical dotted line.}
  \label{fig:fourBody}
\end{figure}

\begin{figure}
  \begin{center} 
    \includegraphics[width=0.7\linewidth]{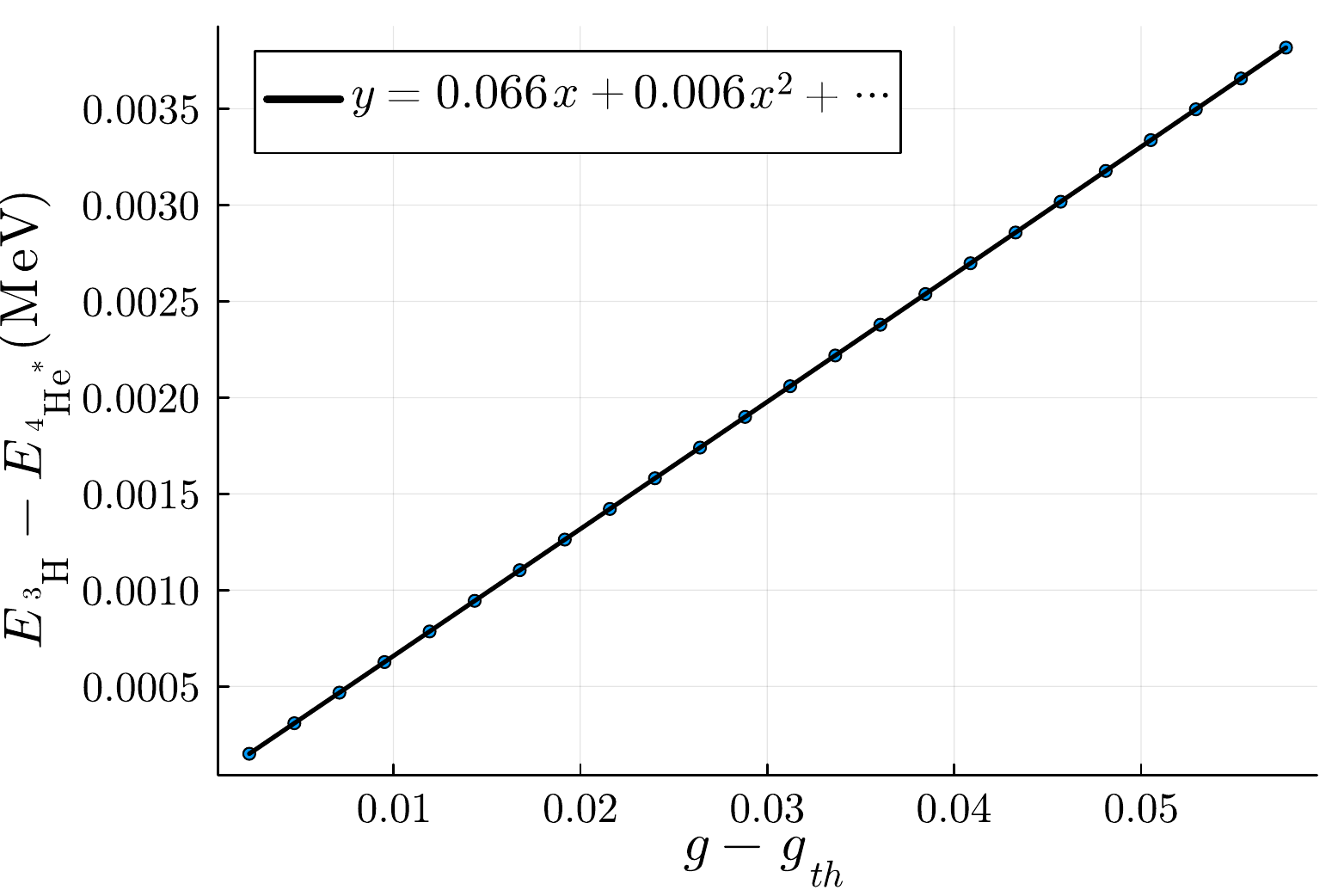}
  \end{center}
  \caption{The energy difference between the four-body excited state and the
  $^3$H+p threshold as a function of $g-g_{th}$. A polynomial fit
  reveals a dominant linear behavior, implying $k(g)\sim i\sqrt{g-g_{th}}$.}
  \label{fig:behaviour4B}
\end{figure}

Another way to track the resonance is by introducing an additional four-body
force that can bind the resonance without altering the three-body thresholds. We choose a
hypercentral four-body force
\begin{equation}
  V_4(\rho_4)  = -g\frac{\hbar^2}{mR_4^2}\,
	e^{- (r_{12}^2+r_{13}^2+r_{14}^2+r_{23}^2+r_{24}^2+r_{34}^2)/R_4^2} \,,
  \label{eq:fourBody}
\end{equation}
with range $R_4=10$~fm, and a coupling constant $g\ge 0$. 
The strength of $g$ is chosen such that the resonant state becomes a bound state.
This occurs for $g > g_{th} = 1.1479$, as shown in Fig.\ref{fig:fourBody}.
Similar to the case of Coulomb interaction, we will use the calculated energies
for $g > g_{th}$ to construct a Pad\'e representation of the momentum $k(g)$ and
analytically continue the momentum up to $g=0$. 
Determining the threshold value is a crucial step in this case too. The
calculations were carried out employing a correlated Gaussian basis set that was
optimized at $g=1.206$ using the stochastic variational
method~\cite{suzuki:1998_}. In Fig.~\ref{fig:behaviour4B} we show 
how the energy difference between the three-body threshold and the 
four-body state behave as a function of function of the the
$g-g_{th}$; even in this case we observe a dominant linear behavior. 

As in the Coulomb case, we use the four-body energies calculated before reaching
the threshold to construct the Pad\'e approximant Eq.(\ref{eq:pade}) for
different values of $N$. The momentum and energy of the resonance are obtained
by extrapolating the approximation to $g=0$. The results are presented in
Table~\ref{tab:resonanceW4}. We observe that the extrapolated value of $E_R =
0.068(1)$ MeV is relatively stable, while the width has a larger uncertainty,
$\Gamma/2 = 0.007(5)$ MeV.
\begin{table}
  \caption{Extrapolated values of the momentum and of the energy of the 
  resonance as a function of the degree of the Pad\'e approximant
  for the four-body interaction.}
  \label{tab:resonanceW4}
\begin{tabular} {@{}c c c c c@{}}
\hline\hline
$N$ & Im($k$)(fm$^{-1}$) & Re(k)(fm$^{-1}$)  & $E_R$ (MeV) & $\Gamma/2$ (MeV)\\ 
\hline
\hline
4.0  & 0.0406 &0.0002  & 0.068 & 0.0006 \\
5.0  & 0.0405 &0.0003  & 0.068 & 0.0010 \\
6.0  & 0.0405 &0.0003  & 0.068 & 0.0010\\
7.0  & 0.0406 &0.0001  & 0.068 & 0.0002\\
8.0  & 0.0403 &0.0003  & 0.067 & 0.0012\\
9.0  & 0.0405 &0.0003  & 0.068 & 0.0008\\
10.0 & 0.0405 &0.0003  & 0.068 & 0.0011\\
\hline
\end{tabular}
\end{table}
%
%
\section{Conclusion}
Some aspects of the spectrum of light nuclei are essentially understood by
the vicinity of the nuclear interaction to the unitary
point~\cite{gattobigio:2019_Phys.Rev.C,kievsky:2014_JournalofPhysics:ConferenceSeries,kievsky:2016_Few-BodySyst,kievsky:2017_Phys.Rev.C},
a property that has been used also to derive a description based on effective
field theory without the degree of freedom of the
pion~\cite{vankolck:1999_Nucl.Phys.A}. 
A class of universality can be identified in which scale invariance,
continuous or discrete, plays a central
role~\cite{gattobigio:2014_Phys.Rev.A,vankolck:2017_Few-BodySyst.,recchia:2022_Few-BodySyst}.
One of the properties shown by four-body systems inside the universal
window is that for each three-body level there are two attached
four-body states~\cite{platter:2004_Phys.Rev.A,gattobigio:2013_Few-BodySyst,gattobigio:2012_Phys.Rev.A}
which are bound with respect to the three-body ground state. They exist also as resonances
attached to higher three-body excited states~\cite{deltuva:2012_Phys.Rev.A}. In fact,
this binary-tree-like structure of two-attached $N+1$-particle state for each
$N$-particle state seems to be a general, but still mysterious,
property of Efimov physics. One suggestive hypothesis is that the pair
of states are just the first two of an infinite tower that in principle
could be exposed by changing a more fundamental four-body
scale~\cite{hadizadeh:2011_Phys.Rev.Lett.,frederico:2023_}.

It is interesting to analyse the spectrum of light nuclear system when
the Coulomb interaction is absent, in this way some universal characterization
of the nuclear levels can be highlighted. As we mentioned, in this case 
there are two four-body bound states, one deep and one shallow attached to the
three-body state, $^3$H, which is degenerate with $^3$He. While the
ground state of the four-body system clearly corresponds to $^4$He, the fate of
the excited state $^4\mathrm{He}^*$ is of interest. By gradually switching on
the Coulomb interaction, the excited state crosses the $^3$H+p threshold, as
shown in Fig.~\ref{fig:threshold}, and likely becomes a resonant state. The
nuclear spectrum exhibits a $0^+$ resonance of $^4$He, raising the question of
whether this resonance is continuously linked to the Efimov excited state. 

To answer this question,
we employed an analytical continuation technique using Pad\'e approximants
\cite{kukulin:1989_} to trace the trajectory of $^4\mathrm{He}^*$ beyond the
threshold. Two different methods were utilized: one involved an analytical
continuation of the Coulomb coupling, and the other employed the introduction of
an artificial four-body force that was gradually released. While the accuracy of
the obtained energies, particularly the resonance width, is not high, the
results provide valuable indications. The Coulomb analysis yielded a central
energy value of $E_R=0.60(3)$ MeV and a width of $\Gamma/2=0.036(6)$ MeV,
whereas the use of the four-body force resulted in $E_R=0.068(1)$ MeV and
$\Gamma/2=0.007(5)$ MeV.

It is indeed intriguing that the obtained results are consistent with
calculations based on ab-initio nuclear interactions, but they deviate from the
accepted values of the $0^+$ resonance energy and width. This discrepancy
highlights the challenges and complexities associated with accurately
determining the properties of resonant states, especially in light nuclei. It
emphasizes the need for further investigations and refined theoretical
approaches to fully understand and reconcile these discrepancies.

\clearpage

\end{document}